\def\year{2020}\relax
\documentclass[letterpaper]{article} % DO NOT CHANGE         
\usepackage{aaai20}  % DO NOT CHANGE THIS
\usepackage{times}  % DO NOT CHANGE THIS
\usepackage{helvet} % DO NOT CHANGE THIS
\usepackage{courier}  % DO NOT CHANGE THIS
\usepackage[hyphens]{url}  % DO NOT CHANGE THIS
\usepackage{graphicx} % DO NOT CHANGE THIS
\usepackage{graphicx}  % DO NOT CHANGE THIS
\usepackage{subcaption}
\usepackage{mwe}
\usepackage{dsfont}
\usepackage{amsmath,amsfonts,amssymb}
\usepackage{caption}
\usepackage{multirow}
\usepackage{booktabs}

\usepackage{bm}
\newcommand{\mb}[1]{\mathbf{#1}}

\DeclareMathOperator*{\relu}{ReLU}

\title{Hotel2vec: Learning Attribute-Aware Hotel Embeddings with Self-Supervision}
\author{Ali Sadeghian,\textsuperscript{\rm 1,2}\thanks{Authors contributed equally} Shervin Minaee,\textsuperscript{\rm 1}$^*$ Ioannis Partalas,\textsuperscript{\rm 1}$^*$ Xinxin Li\textsuperscript{\rm 1} \\ \Large \textbf{ Daisy Zhe Wang,\textsuperscript{\rm 2} Brooke Cowan\textsuperscript{\rm 3}\thanks{Work completed 
while author was at Expedia Group.}}\\ % All authors must be in the same font size and format. Use \Large and \textbf to achieve this result when breaking a line
\textsuperscript{\rm 1} Expedia Group, \, \textsuperscript{\rm 2} University of Florida, \,  \textsuperscript{\rm 3} Oregon Health and Science University\\
\{asadeghian, daisyw\}@ufl.edu, \, \{sminaee, ipartalas, xinxli\}@expedia.com, \, cowanb@ohsu.edu
}

\begin{document}
%\vspace{-3cm}

\maketitle

\thispagestyle{plain}
\pagestyle{plain}

% %File: formatting-instruction.tex
% \documentclass[letterpaper]{article}
% \usepackage{aaai}
% \usepackage{times}
% \usepackage{helvet}
% \usepackage{courier}

% \frenchspacing
% \setlength{\pdfpagewidth}{8.5in}
% \setlength{\pdfpageheight}{11in}
% \pdfinfo{
% /Title (Hotel2vec:  Context Aware Hotel Representation Using a Self-Supervised Learning Approach)
% /Author (Put All Your Authors Here, Separated by Commas)}
% \setcounter{secnumdepth}{0}  
%  \begin{document}
% % The file aaai.sty is the style file for AAAI Press 
% % proceedings, working notes, and technical reports.
% %
% \title{Hotel2vec:  Context Aware Hotel Representation Using a Self-Supervised Learning Approach}
% \author{AAAI Press\\
% Association for the Advancement of Artificial Intelligence\\
% 2275 East Bayshore Road, Suite 160\\
% Palo Alto, California 94303\\
% }

\begin{abstract}
We propose a neural network architecture for learning vector representations of hotels.
Unlike previous works, which typically only use user click information for learning item embeddings, we propose a framework that combines several sources of data, including user clicks, hotel attributes (e.g., property type, star rating, average user rating), amenity information (e.g., the hotel has free Wi-Fi or free breakfast), and geographic information. 
During model training, a joint embedding is learned from all of the above information.
We show that including structured attributes 
about hotels enables us to make better predictions in a downstream task than when we rely exclusively on click data.
We train our embedding model on more than 40 million user click sessions from a leading online travel platform, and learn embeddings for more than one million hotels. 
Our final learned embeddings integrate distinct sub-embeddings 
for user clicks, hotel attributes, and geographic information, 
providing an interpretable representation that can be used flexibly depending on the application.
We show empirically that our model generates high-quality representations that boost the performance of a hotel recommendation system in addition to other applications. 
An important advantage of the proposed neural model is that it addresses the cold-start problem for hotels with insufficient historical click information by incorporating additional hotel attributes which are available for all hotels. 
\end{abstract}

\section{Introduction}

Learning semantic representations (embeddings) of different entities, such as textual, commercial, and physical, has been a recent and active area of research. 
Such representations can facilitate applications that rely on a notion of similarity, for example recommendation systems and ranking algorithms in e-commerce.

In natural language processing, word2vec \cite{word2vec} learns vector representations of words from large quantities of text, where each word is mapped to a $d$-dimensional vector such that semantically similar words have geometrically closer vectors. This is achieved by predicting either the context words appearing in a window around a given target word (skip-gram model), or the target word given the context (CBOW model). 
The main assumption is that words appearing frequently in similar contexts share statistical properties (the \textit{distributional hypothesis}). Crucially, word2vec models, like many other word embedding models, preserve sequential information encoded in text so as to leverage word co-occurrence statistics.
The skip-gram model has been adapted to other domains in order to learn dense representations of items other than words. For example, product embeddings in e-commerce \cite{prod2vec} or vacation rental embeddings in the hospitality domain \cite{airbnb} can be learned by treating purchase histories or user click sequences as sentences and applying a word2vec approach.

Most of the prior work on item embedding exploit the co-occurrence of items in a sequence as the main signal for learning the representation. 
One disadvantage of this approach is that it fails to incorporate rich structured information associated with the embedded items.
For example, in the travel domain, where we seek to embed hotels and other travel-related entities, it could be helpful to encode explicit information such as user ratings, star ratings, hotel amenities, and location in addition to implicit information encoded in the click-stream.

In this work, we propose an algorithm for learning hotel embeddings that combines sequential user click information in a word2vec approach with additional structured information about hotels. 
We propose a neural architecture that adopts and extends the skip-gram model to accommodate arbitrary relevant information of embedded items, including but not limited to geographic information, ratings, and item attributes. 
In experimental results, we show that enhancing the neural network to jointly encode click and supplemental structured information outperforms a skip-gram model that encodes the click information alone. 
The proposed architecture also naturally handles the cold-start problem for hotels with little or no historical clicks. Specifically, we can infer an embedding for these properties by leveraging their supplemental structured metadata.

Compared to previous work on item embeddings, the novel contributions of this paper are as follows:
\begin{itemize}
    \item We propose a novel framework for fusing multiple sources of information about an item (such as user click sequences and item-specific information) to learn item embeddings via self-supervised learning. 
    \item We generate an interpretable embedding which can be decomposed into sub-embeddings for clicks, location, ratings, and attributes, and employed either as separate component embeddings or a single, unified embedding.
    \item It is also dynamic, meaning it is easy to reflect future changes in attributes such as star-rating or addition of amenities in the embedding vectors without retraining. 
    \item We address the cold-start problem by including hotel metadata which are independent of user click-stream interactions and available for all hotels. This helps us to better impute embeddings for sparse items/hotels.
    \item We show significant gains over previous work based on click-embedding in several experimental studies.
\end{itemize} 

The structure of the remainder of this paper is as follows. 
Section 2 gives an overview of some of the recent works on neural embedding. 
Section 3 provides details of the proposed framework, including the neural network architecture, training methodology, and how the cold-start problem is addressed.
In Section 4, we present experimental results on several different tasks and a comparison with previous state-of-the-art work. Section 5 concludes the paper.

\section{Related Work}
\label{sec:prev_work}

Recommendation is an inherently challenging task that requires learning user interests and behaviour. There has been a significant body of research on advancing it using various frameworks ~\cite{pazzani2007content,liu2010personalized,ziesemer2011know,agarwal2013content,guerraoui2017sequences}.
Learning a semantic representation/embedding of the items being recommended is a critical piece of most of these frameworks.

Deep learning models have been widely used for learning embeddings~\cite{wang2018attention,wu2019session,sadeghian2016temporal}.
One prominent use case is learning product embeddings for e-commerce. In~\cite{prod2vec,barkan2016item2vec}, the authors develop an approach based on the skip-gram model~\cite{word2vec}, frequently used in natural language processing. They leverage users' purchase histories obtained from their e-mail receipts to learn a dense representation of products. Each user's complete purchase history is represented as a sequence, which is treated as a sentence in which the items are considered words. 

In more recent work~\cite{airbnb}, the authors use the skip-gram framework to learn embeddings for vacation rental properties. They extend the ideas in~\cite{prod2vec} to take into account a user's click stream data during a session. 
A key contribution of their method is the modification of the skip-gram model to always include the booked hotels in the context of each target token, so that special attention is paid to bookings. 
They also improve negative sampling by sampling from the same market, which leads to better within-market listing similarities. 
Nevertheless, their model relies exclusively on large amounts of historical user engagement data, which is a major drawback when such data are sparse. 

In another relevant work, \cite{youtube}, authors propose a framework for YouTube video recommendation which fuses multiple features (e.g., video watches, search tokens, geo embeddings) into a unified representation via a neural architecture. They then use these embeddings for candidate generation and ranking. 
The main limitation of this work is that the individual embeddings are learned separately, and then combined via a neural network to perform classification.

Similar to our work on hotel2vec, there are also some works which attempt to include explicit item attributes (e.g., size, artist, model, color) within the sequence prediction framework using various strategies. 
In \cite{metaprod2vec}, the item metadata is injected into the model as side information to regularize the item embeddings. In their approach, they only use one feature (singer ID) in the experiments. In addition, their approach does not accommodate learning independent embedding vectors for each attribute group. Most recently, 
\cite{singh} propose a method where they train separate encoders for text data, click-stream session data, and product image data, and then use a simple weighted average to unify these embeddings. The weights are learned using grid search on the downstream task. 
While their approach allows for exploring independent embedding vectors, the sub-embeddings of different attribute groups are learned independently rather than jointly.  In addition to efforts extending the skip-gram framework, emerging research attempts to extend GloVe \cite{pennington2014glove} by incorporating various attributes. 
\cite{jeawack} incorporate attribute information into GloVe by modifying the loss function such that the representation of a location can be learned by combining both text and structural data.

\section{The Proposed Framework}
\label{sec:proposed_framework}

Similar to~\cite{word2vec}, by treating the clicks made by users within an interactive web session as words, and sequences of clicks as sentences, we seek to predict the context hotels (words), given a target hotel (word) in the session (sentence). On a high level, this is the approach proposed in~\cite{prod2vec,airbnb}.
We refer to this approach as a \textit{session-only} model.

As mentioned earlier, one drawback of this approach is that it does not use any information apart from the click data, making it very challenging to make predictions for unseen hotels or hotels with sparse click data. 
In addition, the model may be forced to learn certain semantic features which capture aspects of user interest, hotel geographic information, hotel attributes, and so on, as latent variables as opposed to leveraging them as explicitly-provided input features. 
To address these shortcomings, we propose adding more explicit information about the hotel as model input. Intuitively, this should make the model more efficient during training as well as provide information that it can use when making predictions on unseen or sparse hotels.

Another major advantage of our model is its use of different projection layers for various hotel/item attributes. This enables us to learn independent embedding vectors representing different facets of the property, in addition to an enriched, unified embedding for each hotel.
This model also provides a dynamic framework for updating the embedding of a hotel, once its user-rating or other attribute information changes over time. This is not trivial in session-only models, unless we re-train a new model based on recent click data post attribute changes. In the remainder of the paper, we refer to our proposed model as an \textit{enriched} model, in contrast to the session-only model introduced above.

\subsection{Neural Network Architecture}

\begin{figure}[h!]
    \centering
    \includegraphics[width=0.425\textwidth]{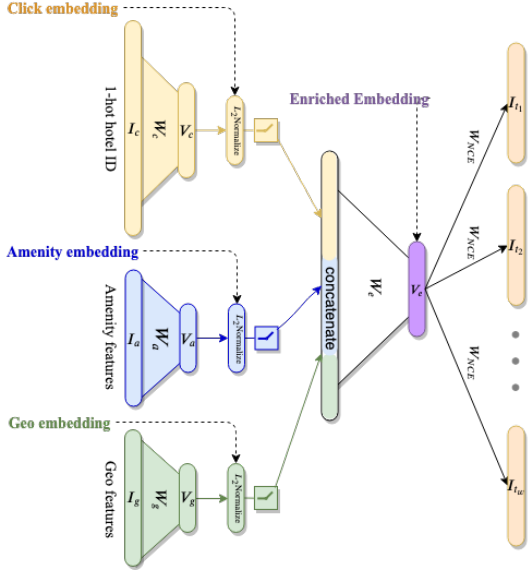}
    \caption{The block-diagram of the enriched, hotel2vec model with a single encoding layer.}
    \label{fig:hotel2vec_model}
\end{figure}

Figure~\ref{fig:hotel2vec_model} illustrates the proposed architecture for an enriched, hotel2vec model.
As we can see, each aspect of the hotel is embedded separately, and these representations are later concatenated and further compressed before being used for context prediction.

Formally, a click session is defined as a sequence of hotels (items) $\{h_1, h_2, \cdots, h_n\}$ clicked on by a user during a defined window of time or visit. We denote the click, amenity, geographic, and enriched embedding vectors with $\mb{V}_c$, $\mb{V}_a$, $\mb{V}_g$, and $\mb{V}_e$ respectively. These are defined as follows:
\begin{align*}
    \mb{V}_c \! &= f(I_c; \mb{W}_c) \\
    \mb{V}_a \! &= f(I_a; \mb{W}_a) \\
    \mb{V}_g \! &= f(I_g; \mb{W}_g)
    % \mb{V}_c \! = \frac{I^\top_c \mb{W}_c}{\| I^\top_c \mb{W}_c \|_{\scriptstyle{2}}}, \,
    % \mb{V}_a \! = \frac{I^\top_a \mb{W}_a}{\| I^\top_a \mb{W}_a \|_{\scriptstyle{2}}}, \,
    % \mb{V}_g \! = \frac{I^\top_g \mb{W}_g}{\| I^\top_g \mb{W}_g \|_{\scriptstyle{2}}} 
\end{align*}
\begin{equation}
 \mb{V}_e = \relu( [\mb{V}_c, \mb{V}_a, \mb{V}_g]^\top \mb{W}_e)
%  f( [\mb{V}_c, \mb{V}_a, \mb{V}_g];  \mb{W}_e)
\end{equation}
where $I_c$ is the one-hot encoding of hotels in the click session, and $I_g$ is a continuous vector with geographical coordinates of the hotel. Amenity features can be categorical or numerical with possible missing values. Thus, $I_a$ is partitioned per feature, where for numerical features we simply use an element of $I_a$ assigned with the value of that feature, and for categorical features with $m$ categories, we assign $m$ elements of $I_a$ and set the corresponding category to $1$ and the others to $0$. If the feature is missing, we set everything to $0$.
$\relu$ is the rectified linear unit activation function~\cite{glorot2011deep} and $f(x; \mb{W})$ is a normalized projection layer parameterized with trainable weights $\mb{W}$, i.e., $f(x; \mb{W}) = \relu(\frac{x \mb{W}}{\hphantom{{}_2}\| x \mb{W} \|_{\scriptstyle{2}}})$.

We train our model using negative sampling based on optimizing the noise contrastive estimation (NCE) loss~\cite{noise_contrastive}. 
More formally, given $h_t$ as the target, we estimate the probability of $h_c$ being a context hotel to be
\begin{equation}
\label{eq:probability_estimate}
\log P(h_c|h_t) = \log \sigma(\mb{V}^\top_{e_{\scriptstyle{t}}} \mb{W}_{c,:})
\end{equation}
where $\mb{W}_{c,:}$ is the $c^{\text{th}}$ row of $W_{\scriptstyle_{NCE}}$. We find parameters of the model by maximizing the probabilities of correct predictions. We train the model using backpropagation and minimizing the following loss function:
\begin{small}
\begin{equation}
\label{eq:training_loss}
% \begin{split}
\! \mathcal{J}(\theta) = - \frac{1}{T} \sum_{t=1}^{T} \biggl( \log P(h_c|h_t) + \!\!\! \sum_{h_i \in \mathcal{N}_c}^{} \!\! {\log \sigma(- \mb{V}^\top_{e_{\scriptstyle{t}}} \mb{W}_{h_i,:}) } \biggr)
% \end{split}
\end{equation}
\end{small}

\noindent where $\mb{V}_{e_{\scriptstyle{t}}}$ is the enriched embedding of $h_t$, $\mb{W}_{i,:}$ is $i^{\text{th}}$ row of $W_{\scriptstyle_{NCE}}$ matrix, $\mathcal{N}_c = \{h_i| 1 \leq i \leq N, h_i \sim P_n(h_c)\}$ is the set of negative examples, and $P_n(h_c)$ is the distribution which we use to pick the negative samples. We train our model by maximizing equation~\ref{eq:training_loss} using batch stochastic gradient descent.

\subsection{Negative Sampling}
\label{sec:neg_sampeling}

It is well known~\cite{noise_contrastive,word2vec,armandpour2019robust} that using negative sampling, a version of noise contrastive estimation, significantly decreases the amount of time required to train a classifier with a large number of possible classes. In the case of recommendation, there is typically a large inventory of items available to recommend to the user, and thus we train our skip-gram model using negative sampling.
However, it is not uncommon that users frequently search exclusively within a particular subdomain. For example, in hotel search, a customer looking to stay in Miami will focus on that market and rarely across different markets. 
This motivates a more targeted strategy when selecting negative samples: we select half of our negative samples following the schema in~\cite{word2vec_2}, i.e., from the complete set of all hotels, and the other half uniformly at random from the same market as the clicked hotel. Throughout this paper, a \textit{market} is defined as a set of similar hotels in the same geographic region. It's worth noting that there may be multiple markets in the same city or other geo region. In the experimental section, we show that this improves the model's within-market similarities and its predictions.

\subsection{Cold Start Problem}
\label{sec:cold_start}

In practice, many hotels/items appear infrequently or never in historical data. Recommender systems typically have difficulty handling these items effectively due to the lack of relevant training data. Apart from the obvious negative impacts on searchability and sales, neglecting these items can introduce a feedback loop. That is, the less these items are recommended, or the more they are recommended in inappropriate circumstances, the more the data reinforces their apparent lack of popularity. 

Dealing with such hotels/items and choosing appropriate weights for them is referred to as the "cold start problem." One of the main advantages of the enriched hotel2vec model over session-only approaches is its ability to better handle cold start cases. Although an item might lack sufficient prior user engagement, there are often other attributes available. For example, in our use case, thousands of new properties are added to the lodging platform's inventory each quarter. While we don't have prior user engagement data from which to learn a click embedding, we do have other attributes such as geographical location, star rating, amenities, etc. Hotel2vec can take advantage of this supplemental information to provide a better cold-start embedding. 

\section{Experimental Results}
In this section, we present several experiments to evaluate the performance of the trained hotel2vec embeddings. 
Before diving into the details of the experiments, we first describe the dataset and model parameters.

\begin{figure*}[t]
    \centering
    \begin{subfigure}[b]{0.49\textwidth}
        \centering
        \includegraphics[width=0.9\textwidth, height=0.7\textwidth]{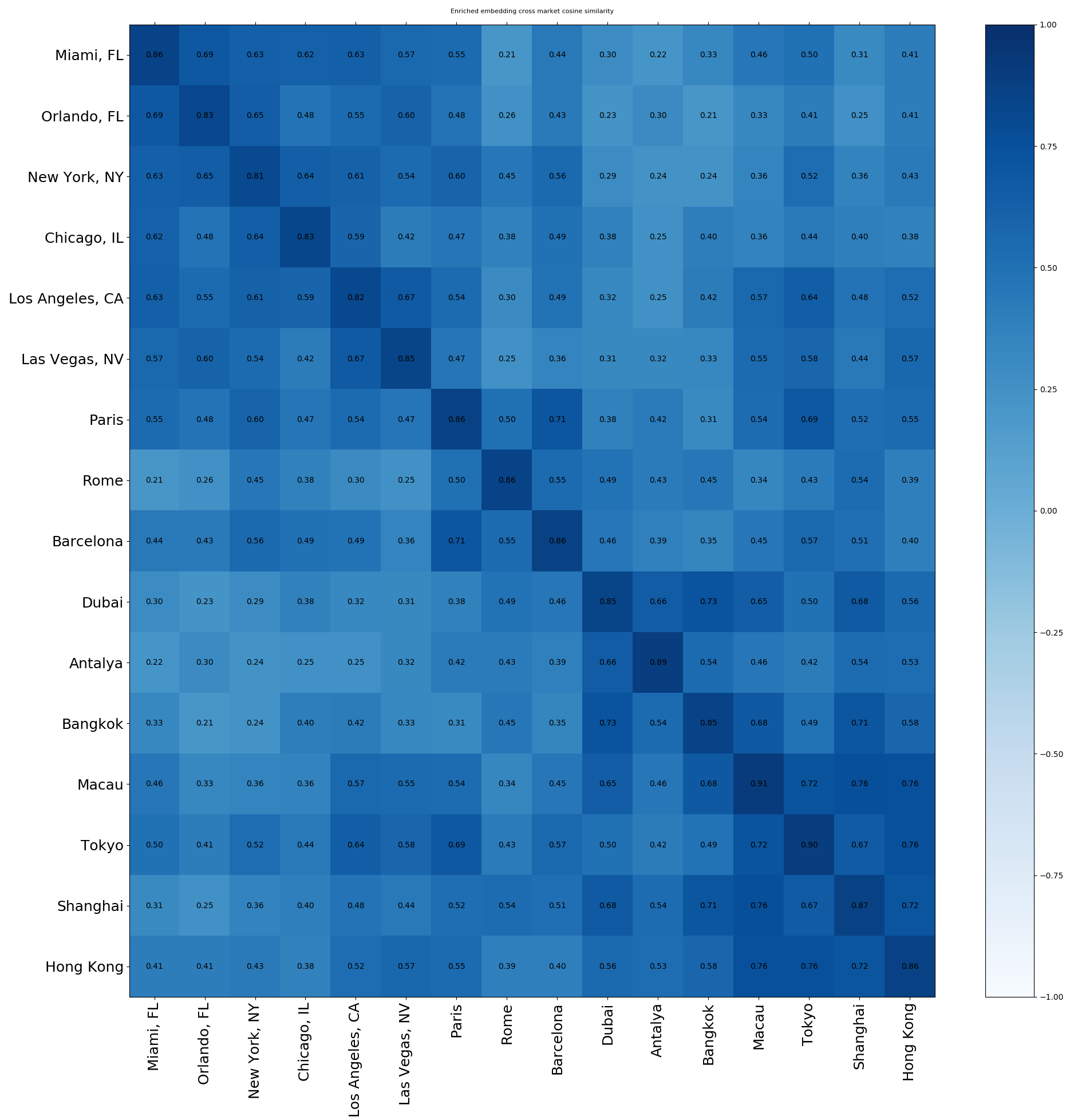}
        \caption[]%
        {{\small Enriched embedding}}    
        \label{fig:enriched_inner_outer_sim_mat}
    \end{subfigure}
    \begin{subfigure}[b]{0.49\textwidth}   
        \centering 
        \includegraphics[width=0.9\textwidth, height=0.7\textwidth]{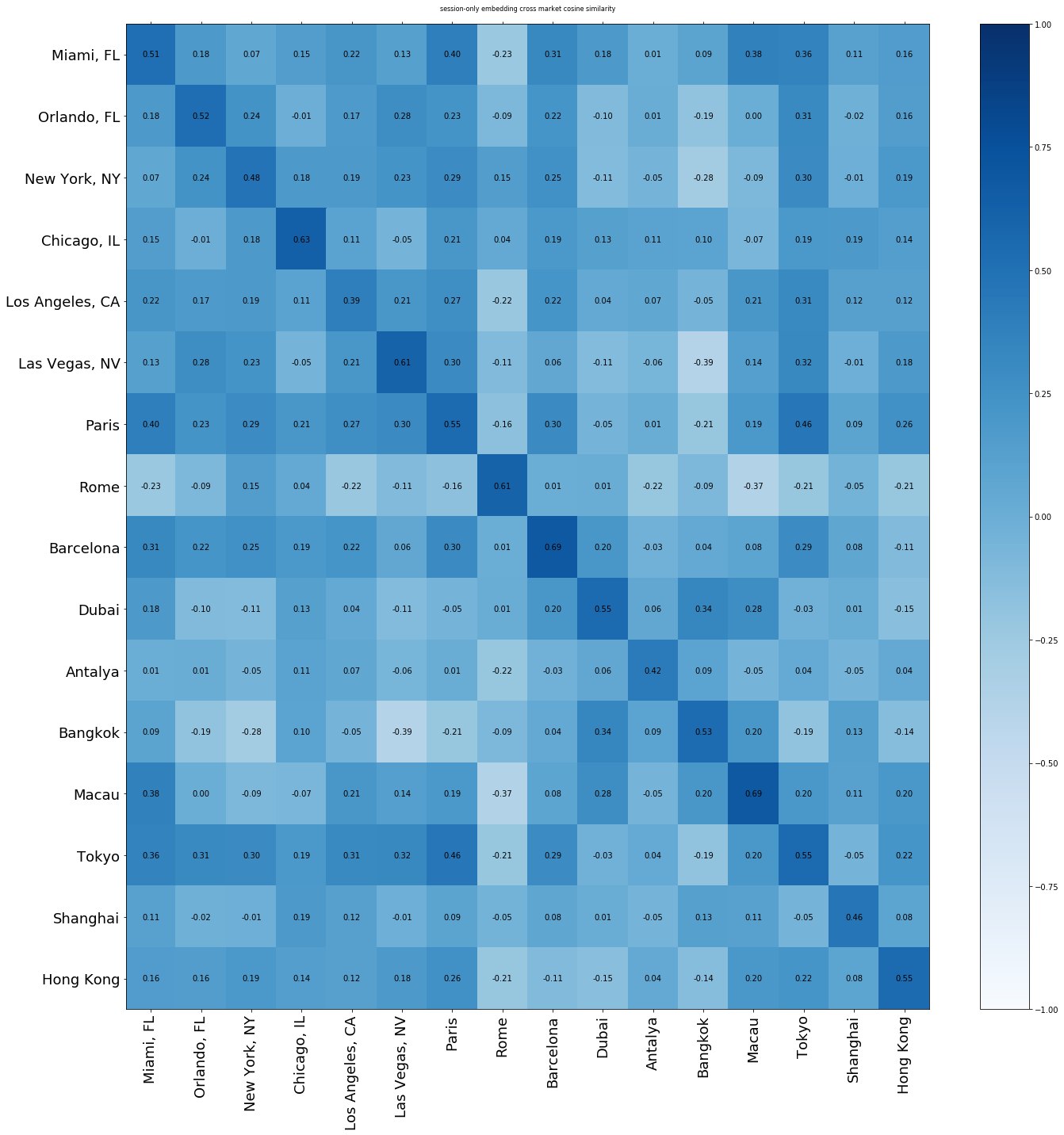}
        \caption[]%
        {{\small Session-only embedding}}    
        \label{fig:session_inner_outer_sim_mat}
    \end{subfigure}%
    % \begin{subfigure}[b]{0.45\textwidth}   
    %     \centering 
    %     \includegraphics[width=0.8\textwidth, height=0.6\textwidth]{"./images/amen_similarity".png}
    %     \caption[]%
    %     {{\small Amenity embedding}}    
    %     \label{fig:amen_inner_outer_sim_mat}
    % \end{subfigure}%
    % \vskip\baselineskip
    % \begin{subfigure}[b]{0.4\textwidth}   
    %     \centering 
    %     \includegraphics[width=\textwidth]{"./images/click_similarity".png}
    %     \caption[]%
    %     {{\small Click embedding}}    
    %     \label{fig:geo_inner_outer_sim_mat}
    % \end{subfigure}
    % \begin{subfigure}[b]{0.4\textwidth}  
    %     \centering 
    %     \includegraphics[width=\textwidth]{"./images/geo_similarity".png}
    %     \caption[]%
    %     {{\small Geo embedding}}    
    %     \label{fig:click_inner_outer_sim_mat}
    % \end{subfigure}
    \caption[]%
    {\small Average cosine similarity of hotels for various  pairs of markets using \textit{enriched} and \textit{amenity} embedding vectors.} 
    \label{fig:market_sim}
\end{figure*}

\subsection{Experimental Framework}
Our dataset contains more than 40M user click sessions, which includes more than 1.1 million unique hotels. 
A click session is defined as a span of clicks performed by a user with no gap of more than 7 days. 
We randomly split the sessions into training, validation, and test with a ratio of 8:1:1.

We use a system with 64GB RAM, 8 CPU cores, and a Tesla V100 GPU. We use Python 3 as the programming language and the Tensorflow~\cite{tensorflow2015-whitepaper} library for the neural network architecture and gradient calculations.
is sufficient for prevention of overfitting.

We tune the hyperparameters for all models, including the baseline session-only model, on the validation set. We search for a learning rate from $\{0.01, 0.1, 0.5, 1.0, 2.5\}$ and embedding dimensions from $\{32, 128\}$. To train the model weights, we use stochastic gradient descent (SGD) with exponential decay since it performs better than other optimizers in our case, and a batch size of 4096. 

For our implementation of the session-only model, a learning rate of 0.5 and embedding dimension of 32 worked best. Throughout the remainder of the paper, we refer to this model as the \textbf{\textit{session-32}} model. For our enriched model (hotel2vec), a learning rate of 0.05 worked best; for the dimensionality of the embedding vectors, we found that letting $V_c, V_e \in \mathds{R}^{32}$, $V_a \in \mathds{R}^{15}$ and $V_g \in \mathds{R}^{5}$ worked best. We refer to this model as the \textbf{\textit{enriched-32}} model.

\subsection{Quantitative Analysis}
\vspace{3pt}

\subsubsection{Hits@k for hotel context prediction}
A robust metric for evaluating a set of hotel embeddings (or, more generally, any set of items displayed to a user in response to an information need) is its ability to predict a user's next click/selection. In this section, we compare our model based on the hits@k metric in various scenarios. Hits@k measures the average number of times the correct selection appears in the top~k~predictions. 

We consider two main scenarios: in the first, we are given the current hotel clicked by the user, and we try to predict the next clicked hotel among all approximately 1.1M hotels (\textbf{\textit{raw}} evaluation). The second scenario is identical except we limit the candidates to hotels within the same market (\textbf{\textit{filtered}} evaluation).

Table~\ref{tab:raw_hits_at_results} shows hits@k for $k \in \{10, 100, 1000\}$ for both the Session-32 and Enriched-32 models. The enriched model outperforms the session-only model by a huge margin, demonstrating the utility of including item attributes when learning embeddings. We also compare both models in the filtered scenario. This is a more realistic case because limiting hotels to the same market reduces the effect of other information the recommender system can use to provide more relevant suggestions to the user.
Table~\ref{tab:filtered_hits_at_results} shows predictions results in the filtered scenario. 

\begin{table}[h!]
\centering
\resizebox{0.9\columnwidth}{!}{%
\begin{subtable}[t]{\linewidth}
\centering
\begin{tabular}{|c|c|c|c|}
\hline
Raw               & \textbf{Hits@10} & \textbf{Hits@100} & \textbf{Hits@1000} \\ \hline
\textbf{Session-32}  & 12.0                & 48.1                 & 72.3              \\ \hline
\textbf{Enriched-32} & 17.4                & 62.2                 & 90.1               \\ \hline
\end{tabular}
\caption{Prediction results of the most likely hotel the user will click next among all possible hotels.}
\label{tab:raw_hits_at_results}
\vspace{10pt}
\begin{tabular}{|c|c|c|c|}
\hline
Filtered          & \textbf{Hits@10} & \textbf{Hits@100} & \textbf{Hits@1000} \\ \hline
\textbf{Session-32}  & 19.6             & 71.7              & 98.1               \\ \hline
\textbf{Enriched-32} & 24.4             & 78.3              & 98.8               \\ \hline
\end{tabular}
\caption{Results when hotel candidates are restricted to the same market as the current hotel.}
\label{tab:filtered_hits_at_results}
\end{subtable}%
}
\caption{Context hotels prediction.}
\label{tbl:main_hits_at_results}
\end{table}

As demonstrated by Table~\ref{tbl:main_hits_at_results}, the enriched model outperforms the baseline session model significantly in both scenarios. This shows the effectiveness of hotel2vec in incorporating both click sessions and item/hotel attributes for better recommendations.

\subsubsection{Comparison using cosine similarity}
In this section, rather than using the model's output probabilities to induce a ranking over hotels, we measure hits@k over the ranking induced using cosine similarity of the embedding vectors. This is useful in scenarios where it isn't feasible to directly use the model's probabilities. 
Table~\ref{tab:cosine_hits_at_results} shows the results for various embeddings. We show that using the enriched vectors one achieves the highest performance.

\begin{table}[h!]
\centering
\resizebox{\columnwidth}{!}{%
\begin{tabular}{|ll|c|c|} 
\hline
\multicolumn{2}{|c|}{Vector used in cosine similarity}    & \textbf{Hits@10}  & \textbf{Hits@100}   \\ 
\hline
\multicolumn{2}{|c|}{\textbf{Session-32} }       & 14.1              & 49.1                \\ 
\hline
\multicolumn{1}{|c}{~~} & {\small \textbf{Click} $(\mb{V}_c)$ }        & 16.7              & 49.5                \\
{\small \textbf{Enriched-32}}  & {\small \textbf{Concatenated} ($[\mb{V}_c, \! \mb{V}_a, \! \mb{V}_g]$)} \hspace{-5pt} & 13.4              & 43.1                \\
                        & {\small \textbf{Enriched} ($\mb{V}_e$) }    & 18.6              & 59.9                \\
\hline
\end{tabular}%
}
\caption{Predicting the next click among all possible hotels using cosine similarity of the vectors.}
\label{tab:cosine_hits_at_results}
\end{table}

We also see from Table~\ref{tab:cosine_hits_at_results} that using cosine similarity instead of the whole network does not result in a huge decrease in performance. Finally, Table~\ref{tab:cosine_hits_at_results} also shows that even the standalone click vectors obtained from the enriched model outperform the embeddings obtained from the session-only model.

\subsubsection{Average intra/inter market embedding similarities}
We expect hotels in the same market to be more similar to each other than to hotels in other markets. To evaluate how well this market-level information is encoded by the learned embeddings, we calculate the average similarity between pairs of markets, with the expectation that we should see a strong diagonal component in the similarity matrix. 
We note that our model is not explicitly trained to learn this kind of market information. However, it is able to learn this by combining the click sessions and hotel attribute information. 
Figure~\ref{fig:market_sim} shows the average similarity scores between hotels in multiple famous cities using two of the embedding vectors. As  Figure~\ref{fig:enriched_inner_outer_sim_mat} clearly depicts, there is a strong similarity between hotels of the same city. Also, markets that are closer to each other (all US cities vs European vs Asian), or for reasons other than geographic proximity are expected to be more similar (e.g., Las Vegas and Macao, or Tokyo and Paris) do indeed have a higher similarity. For comparison, Figure~\ref{fig:session_inner_outer_sim_mat} shows the average cosine similarity between and within markets for the session-only model embeddings.
This model captures within-market similarity well but is not as effective as the enriched model for capturing cross-market similarity. For instance, the session-only model fails to recover the similarity between Las Vegas and Macao.

\subsection{Qualitative Analysis}
\label{sec:exp_hotel_analogy}
\vspace{3pt}

The learned hotel embeddings can be used for recommending similar hotels in various situations. In this section, we show examples of how these embeddings are helpful with real examples of hotels from our dataset. 

\subsubsection{Visualization of embeddings}
\begin{figure*}[h!]
        \centering
        \begin{subfigure}[b]{0.49\textwidth}
            \centering
            \includegraphics[width=0.9\textwidth, height=0.7\textwidth]{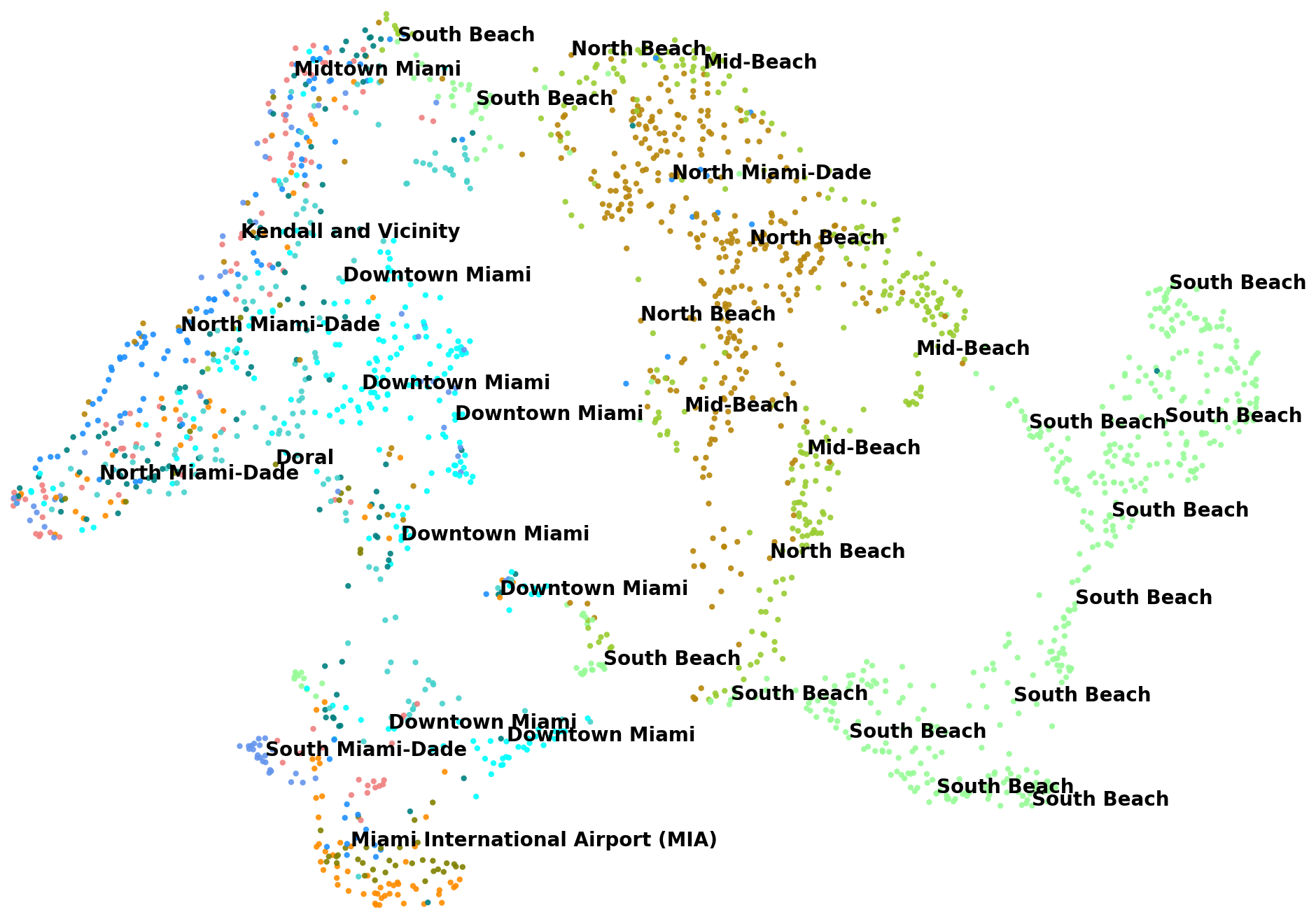}
            \caption[]%
            {{\small Enriched-32 embeddings}}    
            \label{fig:umap_enriched_miami}
        \end{subfigure}
        \begin{subfigure}[b]{0.49\textwidth}   
            \centering 
            \includegraphics[width=0.9\textwidth, height=0.7\textwidth]{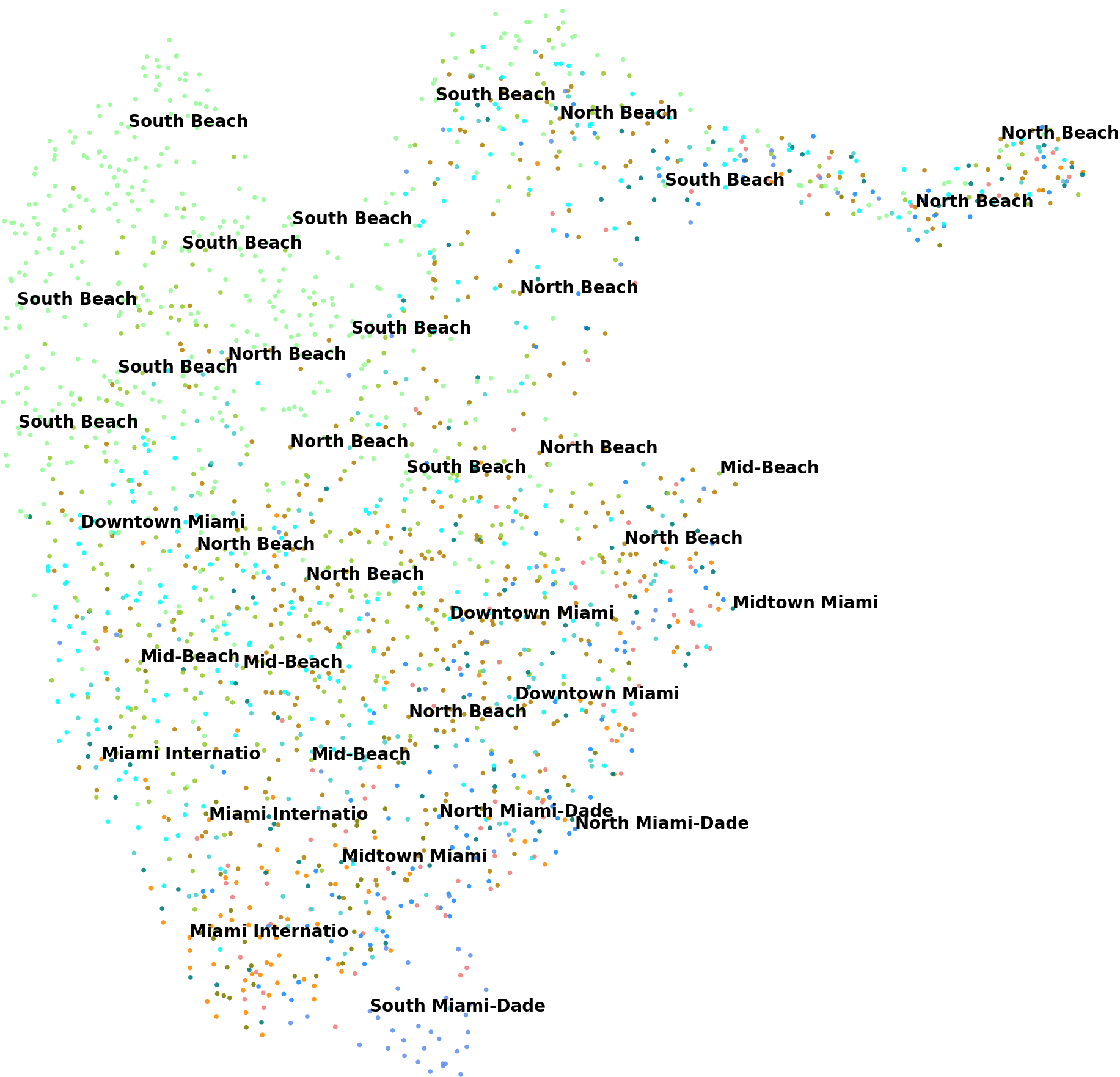}
            \caption[]%
            {{\small Session-32 embeddings}}    
            \label{fig:umap_session_model_miami}
        \end{subfigure}%
        \caption[]%
        {\small Low dimensional visualization of hotel embeddings from the Miami area. Different colors represent expert annotations of competing hotels. Our model has successfully captured most of the similarities.} 
        \label{fig:low_dim_proj}
\end{figure*}

To further illuminate the nature of the embeddings learned by the hotel2vec model, we examine a low-dimensional projection of hotel embeddings in the Miami market (Figures~\ref{fig:umap_session_model_miami}~and~\ref{fig:umap_enriched_miami}). The colors signify the grouping of hotels into various competing subcategories (i.e., similar hotels), manually annotated by a human domain expert. The enriched model is significantly better at clustering similar hotels than the session-only model.

\subsubsection{Most similar hotels}
\begin{figure}[h!]
    \centering
    \includegraphics[width=0.42\textwidth, height=0.40\textwidth]{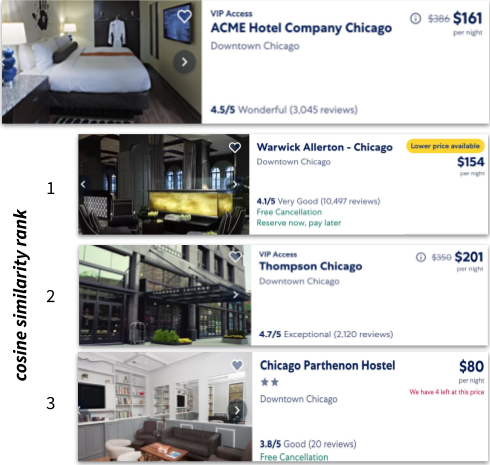}
    \caption{\small Example of recommendations based on cosine similarity of enriched embedding vectors. Ranking by the Session-32 model placed 3\textsuperscript{rd} before 1\textsuperscript{st} (3,1,2), though it is a hostel, cheaper, and has a lower user rating than the target hotel.}
    \label{fig:most_similar_hotels}
\end{figure}

A common scenario is finding similar hotels to a target hotel in other destinations. For example, when the user searches for a specific hotel name (e.g., Hotel Beacon, NY) we would like to be able to recommend a few similar hotels. The learned embeddings can be used to find top-k most similar hotels to a given one. 
Given a target hotel $h$, we compute the cosine similarity of every other hotel with $h$ and pick the most similar hotels. Rigid evaluation of this system requires A/B testing; here we show a few examples comparing our enriched embeddings and the session-only embeddings in Figure~\ref{fig:most_similar_hotels} to provide some intuition for the behavior of the two models.

\subsubsection{Algebraic operations on hotel embeddings}
\begin{figure}[h!]
\hspace{10pt}
    \centering
    \includegraphics[width=0.48\textwidth,height=0.16\textwidth]{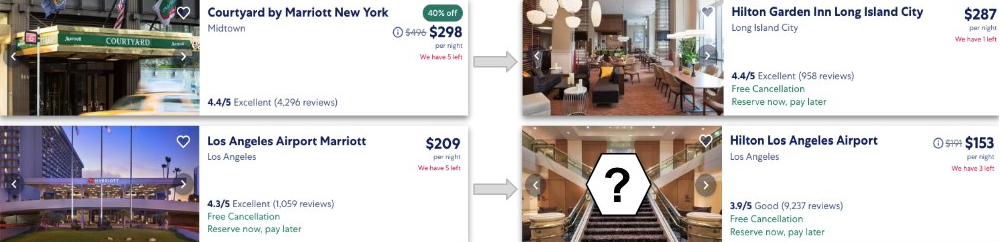}
    \caption{\small Example of algebraic operations on the embeddings for the hotel analogy task.}
    \label{fig:hotel_analogy_example}
\end{figure}

We also investigate whether we can perform meaningful algebraic operations on trained hotel embeddings (similar to the semantic analogy task in~\cite{word2vec}). We pose the question "$h_1$ is to $h_2$ as $h_3$ is to $h_x$" and find $h_x$ as the hotel with the closest vector to $\mb{V_{e_1}}-\mb{V_{e_2}}+\mb{V_{e_3}}$. Figure~\ref{fig:hotel_analogy_example} shows an example of such analogy. $h_1$ is a Marriott hotel in NY, $h_2$ is a Hilton in NY, and $h_3$ is a Marriott in LA (near airport). The obtained $h_x$, is a Hilton hotel in LA near the airport, showing the amount of information captured by the enriched embeddings.

\subsection{Addressing the Cold Start Problem}
\label{sec:exp_cold_start}

Here we analyze how well the model learns embeddings for hotels with little to no presence in the training data. To demonstrate the effectiveness of our model, we compare the enriched model's hits@k with the session-only model's hits@k, for 14K target hotels that were absent during training. Table~\ref{tbl:hits_cold_start_no_imputation} shows results in the filtered scenario. As we can see, the proposed enriched embedding significantly outperforms the session based embeddings for cold-start hotels.

In addition, we use a simple heuristic for cold-start imputation and compare the results with the enriched model for cold-start hotels. To impute vectors for cold-start hotels, we borrow the idea in~\cite{airbnb} and use price, star rating, geodesic distance, type of the property (e.g., hotel, vacation rental, etc.) size in terms of number of rooms, and the geographic market information. For each imputed property, we collect the most similar properties in the same market based on the above features, considering only those properties that fall within a radius of 5km of the target hotel. Results are in Table~\ref{tbl:hits_cold_start_with_imputation}. The heuristic imputation technique improves the Session-32 model's performance on cold-start hotels, but it remains well below that of the enriched model.

\begin{table}[h!]
\centering
\resizebox{0.95\columnwidth}{!}{%
\begin{subtable}[t]{\linewidth}
\centering
\begin{tabular}{|c|c|c|c|}
\hline
Normal   & \textbf{Hits@10} & \textbf{Hits@100} & \textbf{Hits@1000} \\ \hline
\textbf{Session-32}  & 0.21           & 3.68             & 45.01            \\ \hline
\textbf{Enriched-32} & 2.89           & 17.99             & 69.46               \\ \hline
\end{tabular}
\caption{Same-market prediction results when the target hotel is an unseen hotel, using randomly initialized click embeddings.}
\label{tbl:hits_cold_start_no_imputation}
\vspace{10pt}
\begin{tabular}{|c|c|c|c|}
\hline
Imputed   & \textbf{Hits@10} & \textbf{Hits@100} & \textbf{Hits@1000} \\ \hline
\textbf{Session-32}  & 2.70           & 13.15           & 51.73               \\ \hline
\textbf{Enriched-32} & 4.71           & 25.30           & 74.97               \\ \hline
\end{tabular}
\caption{Same-market prediction results when the target hotel is an unseen hotel, click embeddings imputed using 100 hotels in market.}
\label{tbl:hits_cold_start_with_imputation}
\end{subtable}%
}
\caption{Cold start experiments.}
\label{tbl:cold_start_exps}
\end{table}

\subsection{Training Convergence Analysis}
\label{sec:training_convergence}
In this section, we first look at the learning curves for both the session-32 and enriched-32 models. Then, we analyse the effect of $N$ (number of negative samples), $lr$ (learning rate), and the optimization algorithm on the performance of our model. 

\begin{figure}[h!]
    \centering
    \includegraphics[width=0.5\textwidth]{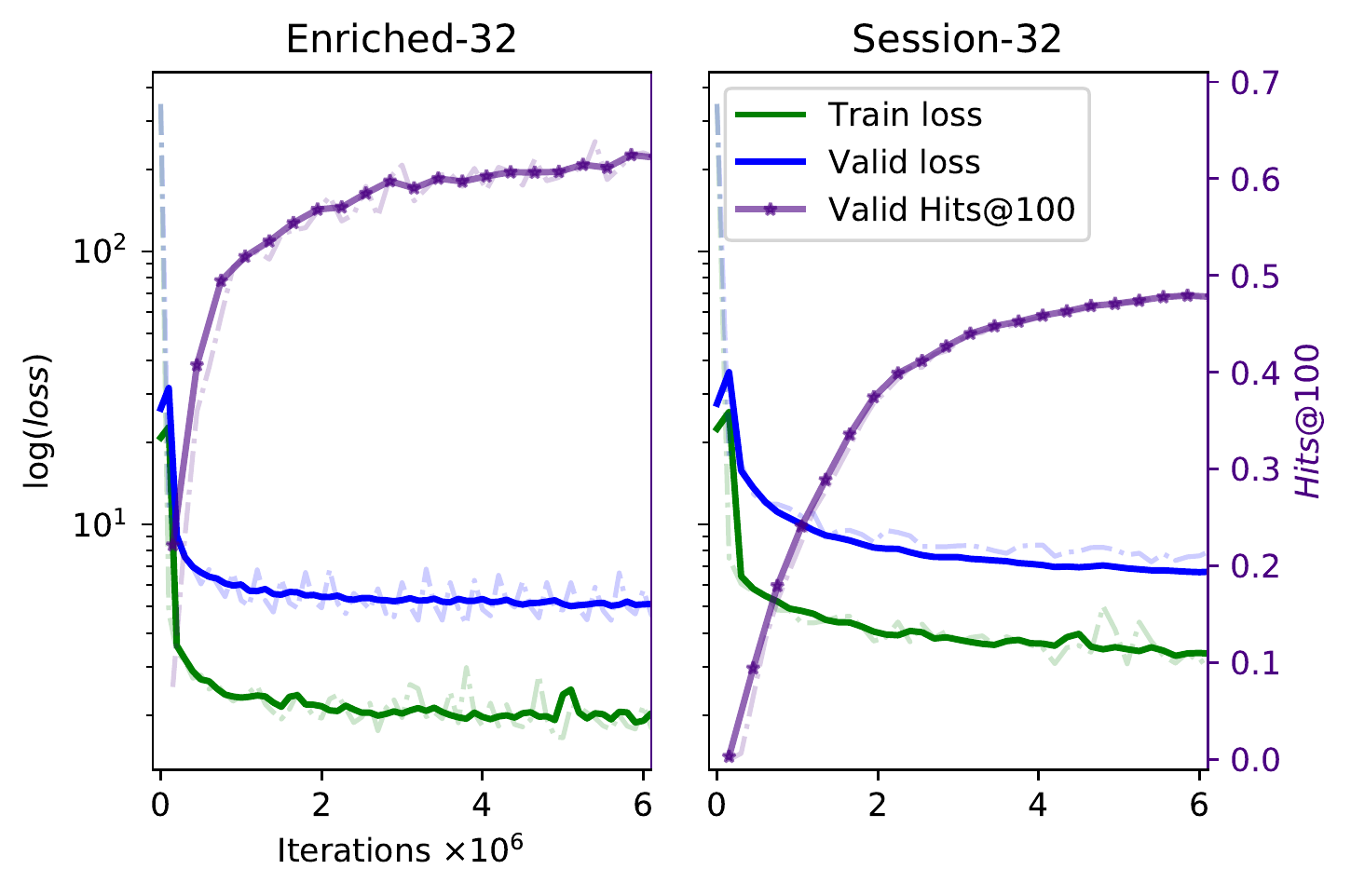}
    \caption{Training progress of both models.}
    \label{fig:training_progress}
\end{figure}

Figure~\ref{fig:training_progress} shows the overall training progress of both the session-32 and enriched-32 models with their respective best hyperparameters. As shown in~Figure~\ref{fig:training_progress}, our model achieves similar performance with fewer data.

\begin{figure}[h!]
    \centering
    \includegraphics[width=0.48\textwidth]{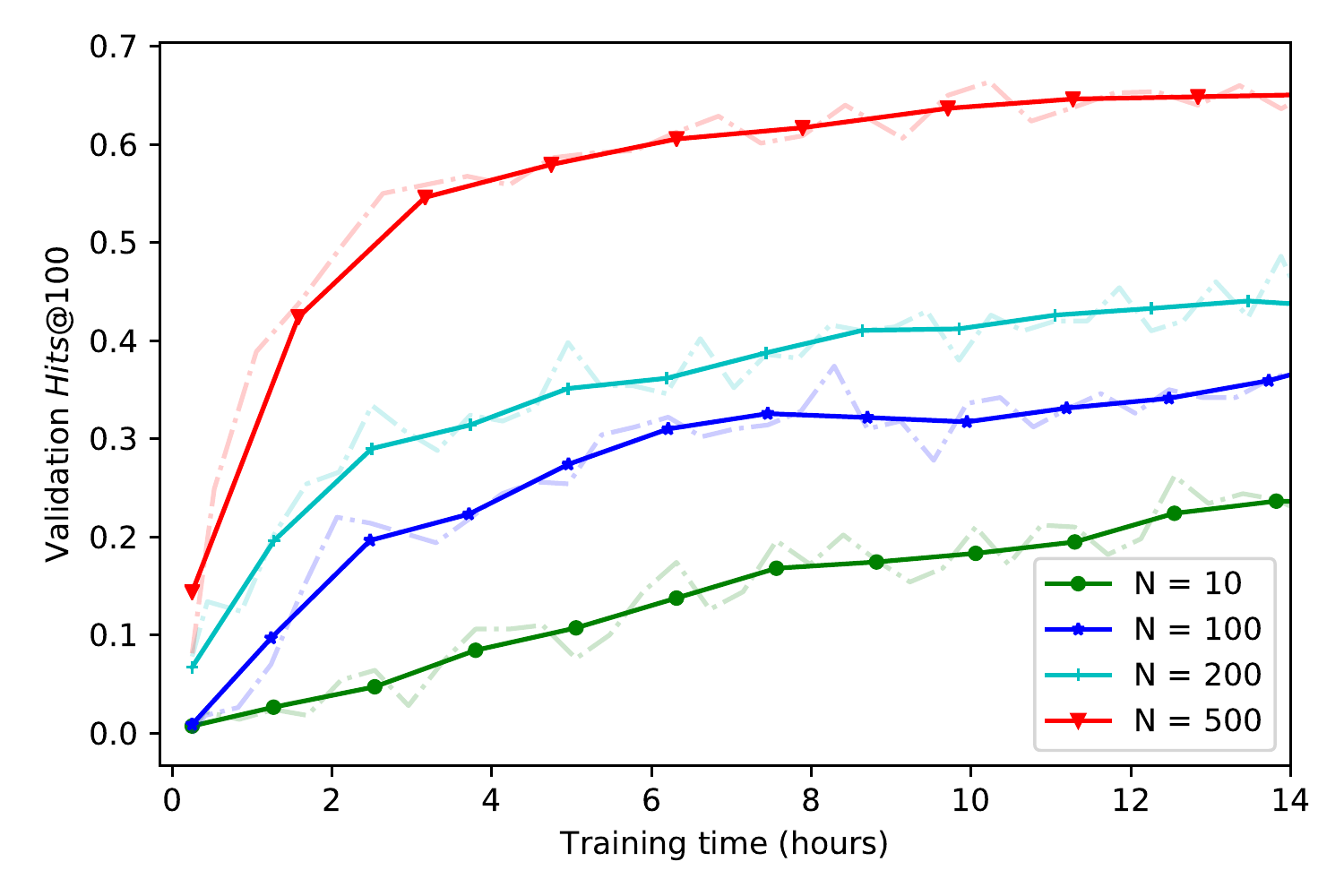}
    \caption{Effect of negative sampling on prediction. Higher number of negative samples results in faster training times.}
    \label{fig:negative_sampling_effect}
\end{figure}

An interesting phenomenon is the effect of increasing the number of negative samples on training time and accuracy. Although it takes more time to create a large number of negative samples, as Figure~\ref{fig:negative_sampling_effect} shows, using more negative samples results in faster training times. 

\begin{figure}[!h]
    \centering
    \includegraphics[width=0.55\textwidth]{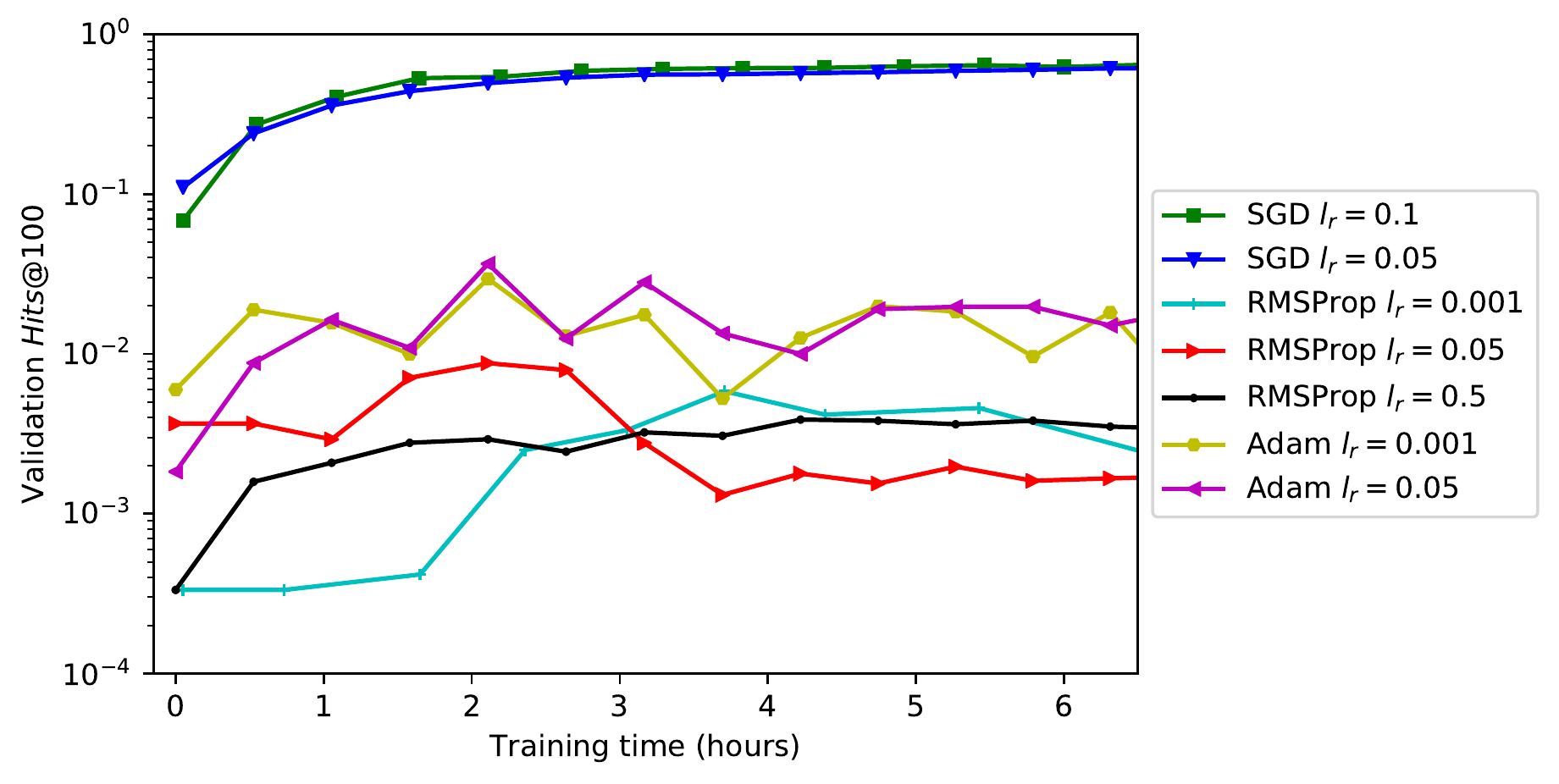}
    \caption{Various optimization algorithms and learning rates. Sophisticated momentum methods seem to overfit to the early batches too quickly.}
    \label{fig:optimizer_effect}
\end{figure}

We show empirical experiments with various optimization algorithms and learning rates, summarized in Figure~\ref{fig:optimizer_effect}. Surprisingly, we see that SGD with exponential learning rate decay outperforms most optimizers with sophisticated learning rate adaptations. We believe this is due to large variance and overfitting in the early stages of training. These issues have been observed in other tasks such as~\cite{popel2018training,devlin2018bert,bashiri2017decomposition}, suggesting the need to use tricks such as warm-up heuristics when using momentum-based optimization algorithms to learn embeddings on large, diverse datasets such as ours.

\section{Conclusion}
In this work, we propose a framework to learn a semantic representation of hotels by jointly embedding hotel click data, geographic information, user rating, and attributes (such as stars, whether it has free breakfast, whether pets are allowed, etc.). 
Our neural network architecture extends the skip-gram model to accommodate multiple features and encode each one separately. 
We then fuse the sub-embeddings to predict hotels in the same session. 
Through experimental results, we show that enriching the neural network with supplemental, structured hotel information results in superior embeddings when compared to a model that relies solely on click information.
Our final embedding can be decomposed into multiple sub-embeddings, each encoding the representation for a different hotel aspect, resulting in an interpretable representation. It is also dynamic, in a sense that if one of the attributes or user ratings changes for a hotel, we can feed the updated data to the model and easily obtain a new embedding. 
Although we mainly focus on learning embeddings for hotels, the same framework can be applied to general item embedding, such as product embedding on Amazon, Ebay, or Spotify.

% No AWK allowed in the blind review version.
\section{ Acknowledgments}
The authors would like to thank Ion Lesan, Peter Barszczewski, Daniele Donghi, Ankur Aggrawal for helping us collecting hotel's attribute, click and geographical data. 
We would also like to thank Dan Friedman and Thomas Mulc for providing useful comments and feedback.

\bibliographystyle{aaai}
\bibliography{sample-base}

\end{document}